\documentclass[prd,aps,showpacs,preprintnumbers,amssymb]{revtex4}

\usepackage{epsf}

\def\e3p{$\eta \rightarrow 3 \pi$}

\begin{document}

\title{%
\hfill{\normalsize\vbox{%
\hbox{\rm SU-4252-810}
 }}\\
{Toy model for two chiral nonets}}                    

\author{Amir H. Fariborz $^{\it \bf a}$~\footnote[3]{Email:
 fariboa@sunyit.edu}}

\author{Renata Jora $^{\it \bf b}$~\footnote[2]{Email:
 cjora@physics.syr.edu}}                    

\author{Joseph Schechter $^{\it \bf 
b}$~\footnote[4]{Email:
 schechte@physics.syr.edu}}

\affiliation{$^ {\bf \it a}$ Department of Mathematics/Science,
 State University of New York Institute of Technology, Utica,
 NY 13504-3050, USA\\\\
$^ {\bf \it b}$ 
Department of Physics,
Syracuse University, Syracuse, NY 13244-1130, USA}
                                                      
\date{\today}

\begin{abstract}

Motivated by the possibility that nonets of scalar mesons
might be described as mixtures of ``two quark" and ``four
quark" components, we further study a toy model in which
corresponding chiral nonets (containing also the
pseudoscalar partners) interact with each other. Although
the ``two quark"  and ``four quark" chiral fields
transform identically under SU(3)$_L \times$ SU(3)$_R$
transformations they transform differently under the
U(1)$_A$ transformation which essentially counts total
(quark + antiquark) content of the mesons. To implement
this we formulate an effective Lagrangian which mocks up
the U(1)$_A$ behavior of the underlying QCD. We derive
generating equations which yield Ward identity type
relations based only on the assumed symmetry structure.
This is applied to the mass spectrum of the low lying
pseudoscalars and scalars. as well as their
``excitations". Assuming isotopic spin invariance, it is
possible to disentangle the amount of``two quark"
vs.``four quark" content in the pseudoscalar $\pi, K
,\eta$ type states and in the scalar $\kappa$ type
states. It is found that a small ``four quark" content in
the lightest pseudoscalars is consistent with a large
``four quark" content in the lightest of the scalar
$\kappa$ mesons. The present toy model also allows one to
easily estimate the strength of a ``four quark" vacuum
condensate. There seems to be a rich and interesting
structure.

\end{abstract}


\maketitle                           

\section{Introduction}

The last few years have seen a renewal of interest \cite{kyotoconf}-
\cite{hss3}
 in the low energy scalar
sector of QCD. Many physicists now believe in the
existence of the light, broad $I=J=0$ resonance, sigma in the
500-600 MeV region as well as a light broad $I,J=1/2,0$ resonance,
kappa in the 700-900 MeV region. Together with the well established
 $f_0(980)$ and $a_0(980)$ scalar resonances, these comprise
a putative nonet of ``elementary particles".
Furthermore, this nonet seems likely to have a quark
structure like $qq{\bar q}{\bar q}$ rather than the conventional
 $q{\bar q}$ \cite{Jaffe}. This of course raises the question
of where are the conventional $q{\bar q}$ p-wave scalars 
expected in the quark model. Arguments have been given
\cite{mixing} that
the experimental data are better fit when the two scalar
nonets mix with each other and the resulting ``level
repulsion", pushes the conventional scalars to higher
masses than otherwise expected.           

    In order to further explore the feature of mixing between
$q{\bar q}$ type and  $qq{\bar q}{\bar q}$ type states it seems 
interesting to consider a linear SU(3)$\times$ SU(3) 
sigma model
which contains also the pseudoscalar nonet partners of these two
scalar nonets. Parenthetically, we remark that while the non-
linear sigma model \cite{GL,CW} and its extension to the chiral 
perturbation theory program \cite{CPT} are 
 often more efficient for systematic calculations,
linear sigma models have a very long history of furnishing
important insights into the nature of strong hadron dynamics.
The SU(2) linear sigma model was first given in ref. \cite{GL}.
It was used as a basis for understanding the current algebra
treatment of $\pi\pi$ scattering
near threshold in ref. \cite{w}. The SU(3) version was given
in the first of ref. \cite{l}. A detailed application to the low energy
pseudoscalar mass spectrum was given \cite{SU1} before QCD in which,
 among other things, it
was shown how a U(1)$_A$ violating term natural in the 
SU(3) model
could solve the $\eta^\prime$ problem. Such a term was later
discovered to arise from instanton effects \cite{tH1}. The 
connection was pointed out in ref. \cite{MS} and emphasized by
't Hooft \cite{tH2}. 

    The model containing two different chiral nonets
 to be discussed here was proposed in section V of
ref. \cite{BFMNS01} and an initial treatment, neglecting flavor
symmetry breaking, was given. A discussion, taking
the flavor symmetry breaking into account has very recently
been presented in ref. \cite{nr}. Actually, it turns out that
the model is very complicated since many different terms can be included
and various assumptions about the nature of the symmetry
breaking can be made. In this paper we will set up
the formalism for treating consequences of the model
which hold (at tree level) just due to the symmetry structure of
the model and will give a numerical treatment using what might be the
simplest choice of symmetry breaking terms.

    Section II begins with a review of the flavor transformation 
properties of the two chiral nonet fields, $M$ and $M'$ which are used 
in the model. Each contains nine pseudoscalar
and nine scalar fields. Under chiral SU(3)$_L\times$ 
SU(3)$_R$ transformations
both fields transform in an identical manner. Thus a chiral Lagrangian
which respects only this symmetry cannot directly distinguish between
a ``two quark" (i.e. $q{\bar q}$) or a ``four quark" scalar, for
example. However, it is noted again that the U(1)$_A$ 
transformation
actually counts the number of quarks in these mesons and provides
a way to distinguish them. In order to make use of this, the Lagrangian
should of course be set up appropriately.
We implement this by requiring that the Lagrangian mock up the
anomalous U(1)$_A$ equation of the underlying QCD and 
that the
analogs of the quark mass terms also mock up the 
 U(1)$_A$ transformation properties of the
quark mass terms in the underlying theory.
 A reasonable
initial thought on which terms to include in the Lagrangian
is to restrict it to be renormalizable. It is noted, with details
in Appendix A, that the renormalizable $M-M'$ Lagrangian has
however
very many more terms than does the renormalizable single $M$
Lagrangian. An alternate way, which still satisfies generality,
is to consider any number of terms, renormalizable or not,
and just use the information which follows from the
symmetry behavior of the Lagrangian.
 
    In order to exploit this symmetry information we derive,
in section III, vector type and axial vector type ``generating
equations" for the model. These can be differentiated with respect to 
the fields to yield many tree level Ward identities which
are independent of the number of symmetric terms included in the 
Lagrangian. In addition to the analog of ``two quark" condensates
which occur in the single $M$ model, the present model also brings 
``four quark" condensates into the picture.

    In section IV, we derive predictions 
for the mass spectrum which follow from this symmetry approach.
The characteristic feature is mixing between``two quark" and
``four quark" mesons with the same quantum numbers. Assuming
isospin invariance, predictions are made for the $\pi-\pi'$
mixing sector, the $K-K'$ mixing sector, the strange
scalar $\kappa-\kappa'$ mixing sector and the sector
involving mixing of the four isocalar pseudoscalars ($\eta$
type particles). It is shown how to formulate the first three
of these mixing sectors in a parallel and economical way.

    In section V, the mass spectrum relations are compared
with experiment. First the three $2\times 2$ mixing sectors
 are treated. The inputs are taken to be the six masses of
the well known and not so well known particles, the pion and kaon
decay constants and a model parameter denoted $x_\pi$, which is
the squared mass of the unmixed (or ``bare") pion. These are enough to 
determine all the relevant parameters of these three systems.
The pseudoscalar mixing is very sensitively dependent on 
$x_\pi$; as it
increases from the experimental value, $m_\pi^2$ the
 four quark components 
of the pion and the kaon increase.   On the other 
hand, the 
scalar $\kappa$ has a large four quark component. This feature
thus provides some support for a more exotic
 structure of the low lying scalars.
Another interesting feature of the present model,
discussed in this section, is that it
permits one to estimate the strength of a
four quark vacuum condensate. Finally, section V contains
a brief summary, the connection with other results
on the same model and directions for future work.

\section{Symmetries and Lagrangian}     
    First, let us briefly review \cite{BFMNS01} the fields of the model 
and their 
transformation properties.
The schematic structure for the matrix $M(x)$ realizing a $q \bar q$ composite
 in terms of quark fields $q_{aA}(x)$ can be written
\begin{equation}
M_a^b = {\left( q_{bA} \right)}^\dagger \gamma_4 \frac{1 + \gamma_5}{2} q_
{aA},
\label{M}
\end{equation}
where $a$ and $A$ are respectively flavor and color indices.  Our convention
 for matrix notation is $M_a^b \rightarrow M_{ab}$.  Then $M$
 transforms under chiral SU(3)$_L \times $ SU(3)$_R$ as
\begin{equation}
M \rightarrow U_L M U_R^\dagger,
\label{Mchiral}
\end{equation}
where $U_L$ and $U_R$ are unitary, unimodular matrices associated with the
 transformations on the left handed
 ($q_L = \frac{1}{2}\left( 1 + \gamma_5 \right) q$) and right
 handed ($q_R = \frac{1}{2}\left( 1 - \gamma_5 \right) q$) quark projections.
  For the discrete transformations charge congugation $C$ and parity $P$
 one verifies 
\begin{equation}
C: \quad M \rightarrow  M^T, \quad \quad P: \quad M({\bf x})
 \rightarrow  M^{\dagger}(-{\bf x}).
\label{MCP}
\end{equation}
 The U(1)$_A$ transformation acts as $q_{aL}
 \rightarrow e^{i\nu} q_{aL}$, $q_{aR} \rightarrow e^{-i\nu} q_{aR}$ and
 results in:
\begin{equation}
M \rightarrow e^{2i\nu} M. 
\label{MU1A}
\end{equation}
Next, consider the $qq{\bar q}{\bar q}$ type fields.
One interesting model \cite{Isgur} 
  postulates that the light scalars are ``molecules''
 made out of two pseudoscalar mesons.  The chiral realization of this
 picture would result in the following schematic structure:
\begin{equation}
M_a^{(2)b} = \epsilon_{acd} \epsilon^{bef}
 {\left( M^{\dagger} \right)}_e^c {\left( M^{\dagger} \right)}_f^d.
\label{M2}
\end{equation}
One can verify that $M^{(2)}$ transforms exactly in the same way
 as $M$ under SU(3)$_L \times$ SU(3)$_R$, $C$ and $P$.
  Under U(1)$_A$ it transforms as 
\begin{equation}
M^{(2)} \rightarrow e^{-4i\nu} M^{(2)},
\end{equation}
which differs from Eq. (\ref{MU1A}).  
Another interesting approach \cite{Jaffe} to explaining the light
scalar mesons was formulated by Jaffe in the framework of the MIT bag
model.  It was observed that the spin-spin (hyperfine) piece of the
one gluon exchange interaction between quarks gives an exceptionally
strong binding to an s-wave $qq\bar q \bar q$ scalar state.
  The scalar states of this
type may be formally written as bound states of a ``dual quark'' and
``dual antiquark''.  There are two possibilities if the dual antiquark
is required to belong to a $\bar 3$ representation of flavor SU(3).
In the first case it belongs to a $\bar 3$ of color and is a spin singlet.
  This has the schematic chiral realization,
\begin{eqnarray}
L^{gE} = \epsilon^{gab} \epsilon^{EAB}q_{aA}^T C^{-1} \frac{1 +
 \gamma_5}{2} q_{bB}, \nonumber \\
R^{gE} = \epsilon^{gab} \epsilon^{EAB}q_{aA}^T C^{-1} \frac{1 - \gamma_5}{2}
 q_{bB}, 
\end{eqnarray}
where $C$ is the charge conjugation matrix of the Dirac theory.  A suitable
 form for the $M$ matrix is:
\begin{equation}
M_g^{(3)f} = {\left( L^{gA}\right)}^\dagger R^{fA}.
\end{equation}
$M^{(3)}$ can be seen to transform in the same way as $M^{(2)}$ under
 SU(3)$_L \times$ SU(3)$_R$, $C$, $P$ and U(1)$_A$.  In 
the second case
 the dual antiquark belongs to a $6$ representation of color and has spin 1.
  It has the corresponding schematic chiral realization:
\begin{eqnarray}
L_{\mu \nu,AB}^g = L_{\mu \nu,BA}^g = \epsilon^{gab} q^T_{aA} C^{-1}
 \sigma_{\mu \nu} \frac{1 + \gamma_5}{2} q_{bB}, \nonumber \\
R_{\mu \nu,AB}^g = R_{\mu \nu,BA}^g = \epsilon^{gab} q^T_{aA} C^{-1}
 \sigma_{\mu \nu} \frac{1 - \gamma_5}{2} q_{bB}, 
\end{eqnarray}
where $\sigma_{\mu \nu} = \frac{1}{2i} \left[ \gamma_\mu, \gamma_\nu \right]
 $.  This choice leads to an $M$ matrix
\begin{equation}
M_g^{(4) f} = {\left( L^{g}_{\mu \nu,AB}\right)}^\dagger R^{f}_{\mu \nu,AB},
\end{equation}
where the dagger operation includes a factor ${(-1)}^{\delta_{\mu 4} +
 \delta_{\nu 4}}$.  $M^{(4)}$ also transforms like $M^{(2)}$ and $M^{(3)}$
 under all of SU(3)$_L \times$ SU(3)$_R$, $C$, $P$ and 
U(1)$_A$.  The 
specific form favored by the MIT bag model calculation actually corresponds to
 a particular linear combination of $M^{(3)}$ and $M^{(4)}$.  Furthermore one
 can verify that $M^{(2)}$ in Eq. (\ref{M2}) is related by a Fierz
 transformation to a linear combination of $M^{(3)}$ and $M^{(4)}$.
  Thus only two of $M^{(2)}$, $M^{(3)}$ and $M^{(4)}$ are linearly
 independent. In any event, at the present effective Lagrangian level,
there are no quantum numbers to distinguish $M^{(2)}$,  $M^{(3)}$,
and  $M^{(4)}$ from each other so we may as well just denote an arbitrary
linear combination of them to be our $qq{\bar q}{\bar q}$ field, 
$M^{\prime}$. Note that $M$ and $M^{\prime}$ are distinguished
from each other by their different U(1)$_A$ 
transformation properties.
These fields may be decomposed into hermitian scalar (S) and pseudoscalar ($\phi$)
nonets as,
\begin{eqnarray}
M &=& S +i\phi, \nonumber  \\
M^\prime &=& S^\prime +i\phi^\prime.
\label{sandphi}
\end{eqnarray}
We will be interested in the situation where non-zero vacuum values
of the diagonal components of $S$ and $S'$ may exist. These will be 
denoted by,
\begin{equation}
\left< S_a^b \right> = \alpha_a \delta_a^b,
 \quad \quad \left< S_a^{\prime b} \right> =
\beta_a \delta_a^b.
\label{vevs}
\end{equation}
In the iso-spin invariant limit, $\alpha_1=\alpha_2$ and
$\beta_1=\beta_2$ while in the SU(3) invariant limit,
$\alpha_1=\alpha_2=\alpha_3$ and $\beta_1=\beta_2=\beta_3$.
                                             
The Lagrangian density which defines our model is
\begin{equation}
{\cal L} = - \frac{1}{2} {\rm Tr}
 \left( \partial_\mu M \partial_\mu M^\dagger
 \right) - \frac{1}{2} {\rm Tr}
 \left( \partial_\mu M^\prime \partial_\mu M^{\prime \dagger} \right)
 - V_0 \left( M, M^\prime \right) - V_{SB},
\label{mixingLsMLag}
\end{equation}
where $V_0(M,M^\prime) $ stands for a general function made
 from SU(3)$_L \times$ SU(3)$_R$ 
(but not necessarily U(1)$_A$) invariants 
formed out of
$M$ and $M^\prime$.  Furthermore $V_{SB}$ is taken to be
a flavor symmetry breaking term which should mock up the quark
mass terms which perform this function in the fundamental
QCD Lagrangian. Other physical particles (including glueballs)
could be added for more realism, but Eq. (\ref{mixingLsMLag}) is already
quite complicated.

    To get an initial indication of what is happening in this kind of 
model the drastically simplified case
where the quark mass effective term, $V_{SB}$ is absent
 and where $V_0$ is simply given by:
\begin{equation}
V_0 = -c_2 {\rm Tr} \left( M M^\dagger \right)
 + c_4 {\rm Tr} \left( M M^\dagger M M^\dagger \right)
 + d_2 {\rm Tr} \left( M^\prime M^{\prime \dagger} \right)
 + e {\rm Tr} \left( M M^{\prime \dagger} + M^\prime M^\dagger \right),
\label{mixingpot}
 \end{equation}
was treated in sec.V of ref. \cite{BFMNS01}.
Here $c_2$, $c_4$ and $d_2$ are positive real constants.  The $M$ matrix
field is chosen to have a wrong sign mass term so that there will be
spontaneous breakdown of chiral symmetry.  A pseudoscalar octet is thus
massless. The mixing between the $M$ and $M^\prime$ is controlled by the 
parameter $e$. The first feature found for this simplified model was that 
 the analog, $\langle{S'}^a_a\rangle$ of the $qq{\bar 
q}{\bar q}$ condensate
in QCD acquired a small non-zero value due to the mixing between
$S$ and $S^\prime$. The main question is the level ordering. Since
the light pseudoscalars (e.g. $\pi^+ =\phi^2_1$) are naturally  
identified, before mixing, with the $q{\bar q}$ field M, one
wonders whether the two quark rather than the four quark scalars
aren't the lightest ones. It was found however that it is
natural (but not unique) in the model to have the energy level pattern
in ascending order-  pseudoscalar Nambu-Goldstone boson  with 
primarily $q{\bar q}$ structure, scalar with primarily $qq{\bar q}{\bar 
q}$ structure, pseudoscalar with primarily $qq{\bar q}{\bar q}$ structure
and scalar with primarily $q{\bar q}$ structure. These refer
to degenerate octets which are each mixtures of $M$ and $M^\prime$
states. This seems to be similar to the expected experimental
pattern and gives us some motivation to proceed further. 

     The next question is what terms to include in the
Lagrangian Eq. (\ref{mixingLsMLag}). A natural first attempt
would be to consider a renormalizable model in which $V_0$ contains
all the SU(3)$\times$ SU(3) invariant terms up to four 
powers of the 
fields. 
These are listed in Appendix A. It is seen that there are 21 terms of 
this 
type. This is a rather large number and while not impossible
to handle suggests trying another tack.  
We will just allow $V_0$ to contain all possible terms which are
SU(3)$_L\times$ SU(3)$_R$ symmetric and use the 
information provided by 
this symmetry. This is more general and also allows for 
non-renormalizable terms. The price to be paid is that we only get 
information which follows just from the symmetry structure.
In an earlier treatment \cite{SU1} of the single chiral nonet case, it was 
found
that the results obtained were essentially those
 which could be obtained from
the ``current algebra" approach.  
 Furthermore, we will try to make use of the fact
that $M$ and $M^\prime$ have different U(1)$_A$ 
transformation properties.
We thus demand that the Lagrangian without $V_{SB}$ mock up
the anomalous U(1)$_A$ equation of QCD,
\begin{equation}
\delta {\cal L} =G,
\label{anomaly}
\end{equation}
where $\delta$ denotes the axial U(1) variation and
 $G$ is proportional to the product of the QCD field strength
tensor and its dual. This can be achieved by making all
of the terms in $V_0$, except for a limited number, 
U(1)$_A$ 
invariant. The special terms will be constructed to satisfy Eq. 
(\ref{anomaly}). An example of a term which is not 
U(1)$_A$
invariant is the mixing term used in 
the simplified model above:
${\rm Tr}\, (M^{\prime}M^{\dagger})+ {\rm h.c.}$. However 
a 
mixing term of the type:
\begin{equation}
\epsilon_{abc}\epsilon^{def}M^a_dM^b_eM^{{\prime}c}_f + 
{\rm h.c.}
\label{invtmixing}
\end{equation}
is U(1)$_A$ invariant and hence possibly the most 
important one.

    An SU(3)$_L\times$ SU(3)$_R$ invariant but not 
U(1)$_A$
invariant term which mocks up Eq. (\ref{anomaly}) can be
seen \cite{U1A} to be 
\begin{equation}
{\cal L}_{anom}= \frac{iG}{12} \,{\rm ln}\,(\frac{{\rm 
det}\, 
M}{{\rm det}\, M^{\dagger}}).
\label{anomterm}
\end{equation}
Here, $G$ is being formally considered as an effective pseudoscalar
glueball field in the effective Lagrangian. To get an $\eta'(958)$
mass term in the effective lagrangian framework one can \cite{U1A}
include a wrong sign mass term for G: $cG^2/2$ in the Lagrangian
which of course does not change the flavor symmetry structure.
Then integrating out $G$ yields the effective  $\eta(960)$
mass term:
\begin{equation}
{\cal L}_\eta =-c_3[{\rm ln}\, (\frac{{\rm det}\, 
M}{{\rm det}\, 
M^{\dagger}})]^2,
\label{etaprmass}
\end{equation}
where $c_3=-1/(288c)$. The nature of this term becomes more
apparent when one goes to the non-linear realization
where $M\rightarrow \alpha_1 {\rm 
exp}\, (i\phi/\alpha_1)$.
For the present paper we shall consider this to be the only
SU(3)$_L\times$ SU(3)$_R$ invariant but not U(1)$_A$
invariant term. However, it is not at all unique when
we consider a model with two chiral nonets. For example
one can also include something like the non- U(1)$_A$
invariant mixing term 
${\rm Tr}\, (M^{\prime}M^{\dagger})+ {\rm h.c.}$ by 
writing
a candidate Lagrangian piece:
\begin{equation}
\frac{iG}{12}[\gamma_1 {\rm ln}\, (\frac{{\rm det}\, 
M}{{\rm det} 
M^{\dagger}})
 +\gamma_2 
{\rm ln} \, 
(\frac{{\rm Tr}\, 
(MM'^{\dagger})}{{\rm Tr}\, (M'M^{\dagger})}],
\label{altanom}
\end{equation}
and proceeding as above. In order to properly mock up
the anomaly in this case it is necessary \cite{hsuss}
that the real numbers $\gamma_1$ and $\gamma_2$ satisfy
\begin{equation}
\gamma_1+\gamma_2=1.
\label{anomnorm}
\end{equation} 
The generalization to more than two such terms is evident. It
may be noted that the $M-M'$ mixing term resulting from
Eq. (\ref{altanom}) mixes only the pseudoscalar fields
and not the scalar ones.

    Finally, let us consider the flavor symmetry breaking terms.
To get more restrictions, we assume that such a term
should mock up both the SU(3)$_L\times$ SU(3)$_R$ and 
U(1)$_A$
transformation properties of the quark mass terms in the fundamental
QCD Lagrangian. It is convenient to introduce a diagonal matrix,
\begin{equation}
A = {\rm diag} (A_1,A_2,A_3),
\label{defineA}
\end{equation}
which is proportional to the diagonal matrix
made from the three
light quark masses, $diag(m_u,m_d,m_s)$ (See \cite{MS} for further 
details). Then, from Eq. (\ref{M}), we note an obvious
 choice for a flavor symmetry breaking term,
\begin{equation}
V_{SB}=-{\rm Tr}\, [A(M+M^{\dagger})]=-2 {\rm Tr}\, (AS),
\label{vsb}
\end{equation}
which transforms like $(3,3^*)+(3^*,3)$ under 
SU(3)$_L\times$ SU(3)$_R$.
Under the U(1)$_A$ transformation of Eq. (\ref{MU1A}), it 
goes to
$-e^{2i\nu} {\rm Tr}\, (AM) + {\rm {\rm h.c.}}$. Note 
that the 
similar 
simple possibility, 
$-2 {\rm Tr} \, (A S')$
does not correctly mock up the U(1)$_A$ transformation
property of the QCD mass term. 
However Eq. (\ref{vsb}) is not at all unique in correctly mocking
up the quark mass term. An interesting term which
does mock up the quark mass term also involves 
mixing and has the form,
\begin{equation}
\epsilon_{abc}\epsilon^{def}A_d^aM_e^b {M'}_f^c + 
{\rm {\rm h.c.}}
\label{sbmixer}
\end{equation}
This term mixes both scalars and pseudoscalars but with
opposite signs.

    For what follows, it is convenient to record the behaviors
of the fields under infinitesimal transformations. Let us write the
infinitesimal vector (L+R) and axial vector (L-R) transformations 
of $\phi$ and $S$ as,
\begin{eqnarray}
\delta_V \phi=[E_V,\phi], \quad \quad \delta_A \phi=-i[E_A,S]_+,
\nonumber \\
\delta_V S=[E_V,S], \quad \quad \delta_A S=i[E_A,\phi]_+.
\label{inftrans}
\end{eqnarray}
Here, unitarity demands that the infinitesimal matrices 
obey,
\begin{equation}
E_V^{\dagger}=-E_V, \quad \quad E_A^{\dagger}=-E_A.
\label{unitarity}
\end{equation}
If we demand that the transformations be unimodular, so
that the U(1)$_A$ transformation is not included
(the U(1)$_V$ transformation is trivial for mesons), we 
should
also impose ${\rm Tr}\, (E_A)=0$. However we will not do 
this so the effects of
U(1)$_A$ will also be included. The transformation 
properties
of the  $qq{\bar q}{\bar q}$ type fields are:
\begin{eqnarray}
\delta_V\phi'=[E_V,\phi'], \quad \quad \delta_A \phi'=-i[E_A,S']_+
+2iS' {\rm Tr} \,(E_A),
\nonumber \\
\delta_V S'=[E_V,S'], \quad \quad \delta_A S'=i[E_A,\phi']_+
-2i\phi' {\rm Tr} (E_A).
\label{inftranspr}
\end{eqnarray}                                                
The extra terms for the axial transformations reflect the different
U(1)$_A$ transformation properties of $M$ and $M'$.

\section{Generating equations}

    We shall consider, in this paper, tree level predictions for the 
Lagrangian of Eq.(\ref{mixingLsMLag}) in which the only 
U(1)$_A$ violating term in $V_0$ is that of 
Eq.(\ref{etaprmass}).
The only term in $V_{SB}$ will be taken to be the simplest one given 
in Eq. (\ref{vsb}). In this minimal picture, there is no symmetry breaking
associated with the $qq{\bar q}{\bar q}$ fields in $M'$. The symmetry
breaking in the  physical states (which contain two
 quark as well as four quark components)
is due to the mixing terms which, as we have
 already seen in Eq. (\ref{invtmixing}),
can be invariant under SU(3)$\times$ SU(3)$\times$ 
U(1)$_A$.
   
 The method of treatment, as used earlier \cite{SU1} to discuss the model
containing only the field $M$, is based on two
generating equations which reflect the invariance of $V_0$ under
vector and axial vector transformations. Differentiating them once,
relates two point vertices (masses) with one point vertices.
 Differentiating them twice relates three point vertices
(trilinear couplings) with masses and so on. These are essentially
tree level Ward identities. 

    Under the infinitesimal vector and axial vector transformations
we have,
\begin{eqnarray}
\delta_VV_0&=&{\{} {\rm Tr}\, (\frac{\partial 
V_0}{\partial \phi}\delta_V \phi
             +\frac{\partial V_0}{\partial S}\delta_VS)
             +(\phi,S)\rightarrow(\phi',S'){\}} =0 ,
    \nonumber \\ 	
\delta_AV_0&=& {\{}{\rm Tr}\, (\frac{\partial 
V_0}{\partial \phi}\delta_A \phi
             +\frac{\partial V_0}{\partial S}\delta_AS) 
             +(\phi,S)\rightarrow(\phi',S'){\}} =- {\cal L}_\eta ,
\label{V0invariance}
\end{eqnarray}
wherein the non-zero value of the axial variation equation  
reflects the presence in $V_0$  of the single $U(1)_A$ non-invariant
term of Eq. (\ref{etaprmass}). Using Eqs. (\ref{inftrans}) and (\ref{inftranspr}) 
as well as the arbitrariness of the variations $E_V$ and $E_A$ yields
 the matrix
 generating equations,
\begin{eqnarray}
&&{\{}[\phi,\frac{\partial V_0}{\partial \phi}]+
[S,\frac{\partial V_0}{\partial S}] + (\phi,S)\rightarrow(\phi',S'){\}} 
=0, 
 \nonumber \\
&&{\{}[\phi,\frac{\partial V_0}{\partial S}]_+ -
[S,\frac{\partial V_0}{\partial \phi}]_+ + 
(\phi,S)\rightarrow(\phi',S'){\}}= 
1[2 {\rm Tr}\, (\phi'\frac{\partial V_0}{\partial S'}-
S'\frac{\partial V_0}{\partial \phi'}) - 
8c_3 i\, {\rm ln}\, (\frac{{\rm det}\, M}{{\rm det}\, 
M^{\dagger}})],
\label{geneqs}
\end{eqnarray}
where, in addition, the form of Eq. (\ref{etaprmass}) was used.
To get constraints on the particle masses we will differentiate
these equations once with respect to each of the four matrix fields:
$\phi,\phi',S,S'$ and evaluate the equations in the ground state.
Thus we also need the ``minimum" condition,
\begin{equation}
\langle\frac{\partial V_0}{\partial S}\rangle + 
\langle\frac{\partial V_{SB}}{\partial S}\rangle=0,
\quad \quad \langle\frac{\partial V_0}{\partial 
S'}\rangle + \langle\frac{\partial V_{SB}}{\partial 
S'}\rangle=0. \label{mincond}
\end{equation}
Using our present choice of Eq. (\ref{vsb}) as the only flavor symmetry breaker
and Eq. (\ref{vevs}), this becomes
\begin{equation}
\langle\frac{\partial V_0}{\partial S_a^a}\rangle = 2A_a, 
\quad \quad 
\langle\frac{\partial V_0}{\partial {S'}_a^a}\rangle = 0. 
\label{firstderiv}
\end{equation}
 Now let us differentiate successively the vector generating equation
 with respect
to $S_a^b$ and to ${S'}_a^b$. This gives with the help of 
Eq.(\ref{firstderiv}),
 the following two relations: 

\begin{eqnarray}
(\alpha_a - \alpha_b)
\langle
{  
  {\partial^2 V_0} \over {\partial S_b^a \partial S_a^b}
}
\rangle
+ (\beta_a - \beta_b)
\langle
{
  {\partial^2 V_0} \over {\partial {S'}_b^a \partial S_a^b}
}
\rangle
&=& 2(A_a-A_b),
   \nonumber \\
(\alpha_a - \alpha_b)
\langle
{
  {\partial^2 V_0} \over {\partial S_b^a \partial {S'}_a^b}
}  
\rangle
+ (\beta_a - \beta_b)
\langle
{
  {\partial^2 V_0} \over {\partial {S'}_b^a \partial{S'}_a^b}
}
\rangle
&=& 0 .
\label{scalarmasses}
\end{eqnarray}
The first of these equations relates the mass mixing transition
 with the 
unprimed scalar squared masses while the second of these relates the
mass mixing transition
 with the primed scalar squared masses. It may be seen
that information is obtained only for particles with different
upper and lower SU(3) tensor indices. In the isospin invariant
limit (where $\alpha_1=\alpha_2$ etc.), information will be
obtained only for the kappa type particles (e.g. $\kappa^+=S_1^3$
when mixing is neglected). If isospin violation information is inserted,
information may be obtained also about the isovector scalars
like $a_0^+(980)$ (which is represented by $S_1^2$ when mixing is 
neglected).
 Next, let us differentiate successively the axial vector generating
 equation with respect to $\phi$ and to $\phi'$. It is neater to write
the results first for the case when fields with different upper and lower
tensor indices are involved: 

\begin{eqnarray}
(\alpha_a + \alpha_b)
\langle
{
  {\partial^2 V_0} \over {\partial \phi_b^a \partial \phi_a^b}
}
\rangle
+ (\beta_a + \beta_b)
\langle
{
  {\partial^2 V_0} \over {\partial {\phi'}_b^a \partial \phi_a^b}
}
\rangle
&=& 2(A_a+A_b),
   \nonumber \\
(\alpha_a + \alpha_b)
\langle
{
  {\partial^2 V_0} \over {\partial {\phi'}_b^a \partial {\phi}_a^b}
}
\rangle
+ (\beta_a + \beta_b)
\langle
{
  {\partial^2 V_0} \over {\partial {\phi'}_b^a \partial {\phi'}_a^b}
}
\rangle
&=& 0
\label{offdiagpsmasses}
\end{eqnarray}

Next, let us write the corresponding equations
 for the case when the upper and lower tensor indices on each field
are the same. 
\begin{eqnarray}
 \alpha_b
\langle
{
  {\partial^2 V_0} \over {\partial \phi_a^a \partial \phi_b^b}
}
\rangle
+ \beta_b
\langle
{
  {\partial^2 V_0} \over {\partial {\phi}_a^a \partial {\phi'}_b^b}
}                                                                          
\rangle
&=& \sum_g \beta_g
\langle
{
  {\partial^2 V_0} \over {\partial {\phi}_a^a \partial {\phi'}_g^g}
}
\rangle
-\frac{8c_3}{\alpha_a}, 
   \nonumber \\
\alpha_b
\langle
{
  {\partial^2 V_0} \over {\partial {\phi'}_a^a \partial \phi_b^b}
}
\rangle
+ \beta_b
\langle
{
  {\partial^2 V_0} \over {\partial {\phi'}_a^a \partial {\phi'}_b^b}
}
\rangle
&=& \sum_g \beta_g
\langle
{
  {\partial^2 V_0} \over {\partial {\phi'}_a^a \partial {\phi'}_g^g}
}                                                                       
\rangle. 
\label{pscalarmasses}
\end{eqnarray}
Note that the axial generating equation provides information on 
the masses of all the pseudoscalars. Further differentiations
will relate a large number of trilinear and quadrilinear
coupling constants to the meson masses and to the quark mass
coefficients, $A_a$.

    To fully characterize the system we will also require some
knowledge of the axial vector and vector currents \cite{SU1} obtained by
Noether's method:
\begin{eqnarray}
(J_\mu^{axial})_a^b &=&(\alpha_a+\alpha_b)\partial_\mu\phi_a^b +
(\beta_a+\beta_b)\partial_\mu{\phi'}_a^b+ \cdots,
\nonumber \\
(J_\mu^{vector})_a^b &=&i(\alpha_a-\alpha_b){\partial_\mu} S_a^b +
i(\beta_a-\beta_b)\partial_\mu {S'}_a^b+ \cdots,
\label{currents}
\end{eqnarray}
where the three dots stand for terms bilinear in the fields.

\section{Predictions for mass spectrum}

    Here we consider the predictions for the mass
spectrum of the model with the Lagrangian given in
 Eq. (\ref{mixingLsMLag}),
whose potential contains any SU(3)$_L\times$ 
SU(3)$_R\times$ U(1)$_A$
invariant terms whatsoever, amended with the 
SU(3)$_L\times$ SU(3)$_R$
but not $U(1)_A$ invariant term of Eq. (\ref{etaprmass}) 
as well as the term, Eq.(\ref{vsb}) 
which transforms exactly like the QCD quark mass term. A
characteristic feature is mixing between fields with the same quantum 
numbers. Specifically, there is information about mixing between
$\pi$ and $\pi'$, between $K$ and $K'$, between $\kappa$
and $\kappa'$ and among among the four $\eta$ type (isosinglet)
states.
 We will take these up in turn. Note that we will be 
working
in the isotopic spin invariant limit \cite{ispinviolation}. 

\subsection{The $\pi-\pi'$ system}

    For compactness let us denote,
\begin{eqnarray}
x_\pi &=& \frac{2A_1}{\alpha_1},
\nonumber \\
y_\pi &=&\langle \frac{\partial^2V}{\partial 
{\phi'}_2^1\partial{\phi'}_1^2}
\rangle,
\nonumber \\
z_\pi&=& \frac{\beta_1}{\alpha_1}.
\label{xyzpi}
\end{eqnarray}
Here we have introduced the total potential $V=V_0+V_{SB}$.
However, since the second derivatives of $V_{SB}$ vanish
with our present choice of flavor symmetry breaker
we may just use $V_0$. Substituting $a=1, b=2$ into
both of Eqs. (\ref{offdiagpsmasses}) enables us to
write the (non-diagonal) matrix of squared $\pi$
and $\pi'$ masses as:
\begin{equation}
(M_\pi^2)=\left[ \begin{array}{c c}
		x_\pi +z_\pi^2 y_\pi & -z_\pi y_\pi
\nonumber \\
		-z_\pi y_\pi   & y_\pi
		\end{array} \right] .
\label{Mpi}
\end{equation}
It is clear that $z_\pi$ is a measure of the mixing
 between $\pi$ and $\pi'$ since the matrix becomes diagonal
 in the limit when $z_\pi$ is set to zero. So we see that
$x_\pi$ would be the squared pion mass in the single M
model and $y_\pi$ represents the squared mass of
the ``bare" $\pi'$. Denoting the eigenvalues of this
matrix by $m_\pi^2$ and $m_{\pi'}^2$, we read off the product 
and sum rules:
\begin{eqnarray}
m_\pi^2m_{\pi'}^2=x_\pi y_\pi,
\nonumber \\
m_\pi^2 + m_{\pi'}^2= x_\pi + y_\pi (1+z_{\pi}^2).
\label{sumrules}
\end{eqnarray}
Assuming that the values of $m_\pi$ and $m_{\pi'}$ are known,
the first of these equations expresses $y_\pi$ in terms of
$x_\pi$. Then the second of these equations 
also expresses $z_{\pi}^2$ in terms of $x_\pi$.
The value of $x_\pi$ is not known but its range
is restricted to be,
\begin{equation}
m_\pi^2 \leq x_\pi \leq m_{\pi'}^2.
\label{xrange}
\end{equation}
This range may be derived by expressing 
 $z_\pi^2$ in terms of $x_\pi$ as mentioned and
requiring $z_\pi^2\geq0$.

    The transformation between the diagonal
 fields (say $\pi^+$ and $\pi'^+$)  and the
original pion fields is defined as:
\begin{equation}
\left[
\begin{array}{c}  \pi^+ \\
		 \pi'^+
\end{array}
\right]
=
\left[
\begin{array}{c c}	
		{\rm cos}\, \theta_\pi & -{\rm sin} \, 
\theta_\pi
\nonumber		\\
		{\rm sin} \, \theta_\pi & {\rm cos} \, 
\theta_\pi
\end{array}		
\right]
\left[
\begin{array}{c}
			\phi_1^2 \\
			{\phi'}_1^2
\end{array}
\right].
\label{mixingangle}
\end{equation}
The explicit diagonalization gives an expression for the 
mixing angle $\theta_\pi$:
\begin{equation}
{\rm tan}\, (2\theta_\pi)=\frac{-2y_\pi 
z_\pi}{y_\pi(1-z_\pi^2)-x_\pi},
\label{thetasubpi}
\end{equation}
which evidently is also known, up to a sign choice for 
$z_\pi$, once $x_\pi$ is specified.

 The mixing angle, $\theta_\pi$ can also be connected to the 
experimentally known value of the pion decay constant (i.e.
the amplitude for the $\pi^+$ meson to decay to two leptons).
Substituting the expressions from Eq. (\ref{mixingangle}) for $\phi_1^2$
and ${\phi'}_1^2$ in terms of the physical fields $\pi^+$
and $\pi'^+$ into Eq. (\ref{currents}) yields,
\begin{eqnarray}
(J_\mu^{axial})_1^2 &=&F_\pi\partial_\mu \pi^+ + F_{\pi'}\partial_\mu 
\pi'^+
+\cdots,
\nonumber \\
F_\pi &=&(\alpha_1+\alpha_2){\rm cos}\, \theta_\pi - 
(\beta_1+\beta_2){\rm sin}\, \theta_\pi,
\nonumber \\
F_{\pi'} &=&(\alpha_1+\alpha_2){\rm sin}\, \theta_\pi + 
(\beta_1+\beta_2){\rm cos}\, \theta_\pi.
\label{Fpis}
\end{eqnarray}
We can then obtain $\alpha_1$ (in the isospin invariant limit) as,
\begin{equation}
\alpha_1=\frac{F_\pi}{2({\rm cos}\, \theta_\pi-z_\pi 
{\rm sin}\, \theta_\pi)}.
\label{alpha1fromF}
\end{equation}
We then successively obtain $A_1$ from the definition of $x_\pi$,
Eq. (\ref{xyzpi}) and $\beta_1$ from the definition of $z_\pi$,
 Eq, (\ref{xyzpi}).
To sum up, specifying $x_\pi$ and the experimental quantities
$m_\pi, m_{\pi'}$ and $F_\pi$ determines all the other parameters of the 
$\pi-\pi'$ system.

\subsection{The $K-K'$ system}

The treatment of this system is almost exactly analogous to that
of the $\pi-\pi'$ system above when one defines the
analogous variables,
\begin{eqnarray}
x_K &=& \frac{2(A_3+A_1)}{\alpha_3+\alpha_1},
\nonumber \\
y_K &=&\langle \frac{\partial^2V}{\partial{\phi'}_3^1\partial{\phi'}_1^3}
\rangle,
\nonumber \\  
z_K&=& \frac{\beta_3+\beta_1}{\alpha_3+\alpha_1}.
\label{xyK}
\end{eqnarray}
 Substituting $a=1, b=3$ into
both of Eqs. (\ref{offdiagpsmasses}) enables us to
write the (non-diagonal) matrix of squared $K$
and $K'$ masses as:
\begin{equation}
(M_K^2)=\left[ \begin{array}{c c}
                x_K +z_{K}^2 y_K & -z_K y_K
\nonumber \\
                -z_K y_K   & y_K
                \end{array} \right] .
\label{MK}
\end{equation}                                                         
This is observed to be identical to the expression for $(M_{\pi}^2)$
in Eq. (\ref{Mpi}) when one simply substitutes everywhere $K$ for $\pi$
and $K'$ for $\pi'$. Similarly, the four equations (\ref{sumrules}),
 (\ref{xrange}), (\ref{mixingangle}) and (\ref{thetasubpi}) continue to 
hold when one substitutes everywhere  $K$ for $\pi$ and  $K'$ for $\pi'$.
Similarly, the $K^+$ decay constant, $F_K$ is now defined from,
\begin{eqnarray}
(J_\mu^{axial})_1^3 &=&F_K\partial_\mu K^+ + F_{K'}\partial_\mu
K'^+
+\cdots,
\nonumber \\
F_K &=&(\alpha_1+\alpha_3){\rm cos}\, \theta_K -
(\beta_1+\beta_3){\rm sin}\, \theta_K,
\nonumber \\
F_{K'} &=&(\alpha_1+\alpha_3){\rm sin}\, \theta_K +
(\beta_1+\beta_3){\rm cos}\, \theta_K.
\label{FKs}.
\end{eqnarray}                                                           

We can then obtain $\alpha_3 +\alpha_1$ (in the isospin invariant limit) 
as,
\begin{equation}
\alpha_3+\alpha_1=\frac{F_K}{{\rm cos}\, \theta_K-z_K 
{\rm sin}\, \theta_K}.
\label{strangealph}
\end{equation}
We then successively obtain $A_3+A_1$ from the definition of $x_K$
 and $\beta_3 + \beta_1$ from the definition of $z_K$.
To sum up, specifying $x_K$ and the experimental quantities
$m_K, m_{K'}$ and $F_K$ determines all the other parameters of the
$K-K'$ system.                                                          

\subsection{The $\kappa-\kappa'$ system}
 
Again, we can treat this system in an exactly analogous way to
the $\pi-\pi'$ and $K-K'$ cases if we define the analogous 
quantities:
\begin{eqnarray}
x_\kappa &=& \frac{2(A_3-A_1)}{\alpha_3-\alpha_1},
\nonumber \\
y_\kappa &=&\langle \frac{\partial^2V}{\partial {S'}_3^1\partial {S'}_1^3}
\rangle,
\nonumber \\  
z_\kappa &=& \frac{\beta_3-\beta_1}{\alpha_3-\alpha_1}.
\label{xyzkappa}
\end{eqnarray}
In this case, however, the vector generating equations in Eqs. 
(\ref{scalarmasses}) with the choices $a=1$ and $b=3$ are used.
The transformation between the diagonal and original strange
scalar fields is given by,
\begin{equation}
\left[
\begin{array}{c}  \kappa^+ \\
                 \kappa'^+
\end{array}
\right]
=
\left[
\begin{array}{c c}
                {\rm cos} \, \theta_\kappa & -{\rm sin}\, 
\theta_\kappa
\nonumber               \\                             
         {\rm sin}\, \theta_\kappa & {\rm cos}\,  
\theta_\kappa
\end{array}
\right]
\left[
\begin{array}{c}
                        S_1^3 \\
                        {S'}_1^3
\end{array}
\right],
\label{kappamixingangle}                                         
\end{equation}
where the mixing angle is determined by the diagonalization:
\begin{equation}
tan(2\theta_\kappa)=\frac{-2y_\kappa z_\kappa}
{y_\kappa(1-z_\kappa^2)-x_\kappa}.
\label{thetasubkappa}
\end{equation}                                               
We may define $\kappa$ ``decay constants' as,
\begin{eqnarray}
F_\kappa &=&(\alpha_3-\alpha_1){\rm cos}\, \theta_\kappa 
-
(\beta_3-\beta_1){\rm sin}\, \theta_\kappa,
\nonumber \\
F_{\kappa'} &=&(\alpha_3-\alpha_1) {\rm 
sin}\, \theta_\kappa +
(\beta_3-\beta_1){\rm cos}\, \theta_\kappa,
\label{Fkappas}
\end{eqnarray}                                      
although there is no direct experimental information 
available about them.
 
    Now let us consider the $\pi-\pi'$, $K-K'$ and
$\kappa-\kappa'$ systems together. Using the first two
we can get all of $A_1, A_3, \alpha_1,\alpha_3, \beta_1,
\beta_3$ from the experimental masses of $\pi,\pi',K,K'$,
the experimental decay constants $F_\pi,F_K$
and the assumed values of $x_\pi$ and $x_K$, as seen above.
This means that $x_\kappa$ and $z_\kappa$ may be read off
directly from Eqs. (\ref{xyzkappa}) while $y_\kappa$
can be found from the 
product rule $m_\kappa^2m_{\kappa'}^2=x_\kappa y_\kappa$
if $m_\kappa$ and $m_{\kappa'}$ are furnished. Thus all
the parameters of the $\kappa-\kappa'$ system are known,
given the input masses and the values of $x_\pi$ and
$x_K$. However we have not yet made use of the sum rule
analogous to the second of Eqs. (\ref{sumrules}). This
 provides another way to calculate $z_\kappa$ so we get the consistency
condition: 
\begin{equation}    
(\frac{\beta_3-\beta_1}{\alpha_3-\alpha_1})^2= \frac{x_\kappa(m_\kappa^2
+m_{\kappa'}^2-x_\kappa)}{m_\kappa^2m_{\kappa'}^2} -1
\label{consistentcond}
\end{equation}
Since the quantities in this equation depend on both $x_\pi$ and $x_K$,
the solution can determine the value of $x_K$ for each choice of $x_\pi$.
 In other words, if $x_\pi$
is specified, the parameters of the $\pi-\pi'$, the $K-K'$ and
the $\kappa-\kappa'$ systems are all determined in the present model.

\subsection{The $\eta$ system}

    This system is more complicated because, even in the 
isotopic spin invariant limit, there are four different $I=0$
pseudoscalars which can mix with each other. These may be put together
as a column vector according to,
\begin{equation}
\Phi_0 =
\left[
\begin{array}{c}
\frac{\phi_1^1+\phi_2^2}{\sqrt{2}} \\
\phi^3_3\\
\frac{{\phi'}_1^1+{\phi'}_2^2}{\sqrt{2}}  \\
{\phi'}^3_3
\end{array}
\right].
\label{etabasis}
\end{equation}                                 
The part of the Lagrangian describing the masses of the $I=0$
pseudoscalars is then:
${\cal L} = -(1/2) \Phi_0^T (M^2_\eta) \Phi_0$, where $(M^2_\eta)$
is a symmetric $4\times4$ matrix. Relations among the matrix elements
follow by using both of Eqs. (\ref{pscalarmasses}). These connect
the transition masses both to the ``bare" unprimed particle masses
and to the ``bare" primed particle masses. The use of isospin
invariance relations like the ones given in Appendix B may also be useful.
Eventually, the matrix elements of  $(M^2_\eta)$ depend on 
four new quantities in addition to the ones appearing in the above
three subsystems. The resulting matrix elements are listed below:

\begin{eqnarray}
\left( M^2_\eta \right)_{11} &=& 
{ {2 A_1} \over \alpha_1} - { {16 c_3}\over \alpha_1^2 }
-  { {\beta_1^2 m_\pi^2 m_{\pi'}^2} \over {2 A_1 
\alpha_1} }
+ 2 \left( {\beta_1 \over \alpha_1 } \right)^2
     \langle { {\partial^2 { V}}
                        \over
               {(\partial {\phi'}^1_1)^2}
             }
     \rangle
+    4 \left( { {\beta_1\beta_3} \over {\alpha_1^2} } 
\right)
     \langle { {\partial^2 { V} }
                        \over
               {\partial {\phi'}^1_1 \partial {\phi'}^3_3}
             }
     \rangle
\nonumber \\ 
&&+ 2 \left( {\beta_3\over \alpha_1} \right)^2
     \langle { {\partial^2 { V} }
                        \over
               {\partial {\phi'}^3_3 \partial 
{\phi'}^3_3}
             }
     \rangle
\nonumber \\ 
\left( M^2_\eta \right)_{12} &=& 
- { {8 \sqrt{2} c_3} \over {\alpha_1 \alpha_3} }
-  { {\beta_1^2 m_\pi^2 m_{\pi'}^2} \over {\sqrt{2} A_1 
\alpha_3} }
+ \left( 
{  {2\sqrt{2} \beta_1^2} \over {\alpha_1 \alpha_3} } 
  \right)
     \langle { {\partial^2 { V} }
                        \over
               {(\partial {\phi'}^1_1)^2}
             }
     \rangle
+    \left( 
{ {2\sqrt{2}\beta_1\beta_3} 
\over 
{\alpha_1\alpha_3} } 
\right)
     \langle { {\partial^2 { V} }
                        \over
               {\partial {\phi'}^1_1 \partial {\phi'}^3_3}
             }
     \rangle
\nonumber \\ 
\left( M^2_\eta \right)_{13} &=& 
-  { {\beta_1 m_\pi^2 m_{\pi'}^2} \over {2 A_1 } }
+ 2 \left( {\beta_1 \over \alpha_1 } \right)
     \langle { {\partial^2 { V} }
                        \over
               {(\partial {\phi'}^1_1)^2}
             }
     \rangle
+    2 \left( { {\beta_3} \over {\alpha_1} } 
\right)
     \langle { {\partial^2 { V} }
                        \over
               {\partial {\phi'}^1_1 \partial {\phi'}^3_3}
             }
     \rangle
\nonumber \\ 
\left( M^2_\eta \right)_{14} &=& 
\left( {\sqrt{2}\beta_1 \over \alpha_1 } \right)
     \langle { {\partial^2 { V} }
                        \over
               {\partial {\phi'}^1_1 \partial {\phi'}^3_3}
             }
     \rangle
+     \left( { {\sqrt{2}\beta_3} \over {\alpha_1} } 
\right)
     \langle { {\partial^2 { V} }
                        \over
               {\partial {\phi'}^3_3 \partial 
{\phi'}^3_3}
             }
     \rangle
\nonumber \\ 
\left( M^2_\eta \right)_{22} &=& 
{ {2 A_3} \over \alpha_3} - { {8 c_3}\over \alpha_3^2 }
-  { {\alpha_1\beta_1^2 m_\pi^2 m_{\pi'}^2} \over {A_1 
\alpha_3^2} }
+ 4 \left( {\beta_1 \over \alpha_3 } \right)^2
     \langle { {\partial^2 { V} }
                        \over
               {(\partial {\phi'}^1_1)^2}
             }
     \rangle
\nonumber \\ 
\left( M^2_\eta \right)_{23} &=& 
-  { {\alpha_1\beta_1 m_\pi^2 m_{\pi'}^2} \over {\sqrt{2} 
A_1 \alpha_3} }
+ \left( 
{  {2\sqrt{2} \beta_1} \over {\alpha_3} } 
  \right)
     \langle { {\partial^2 { V} }
                        \over
               {(\partial {\phi'}^1_1)^2}
             }
     \rangle
\nonumber\\
\left( M^2_\eta \right)_{24} &=& 
\left( { {2\beta_1} \over {\alpha_3} } 
\right)
     \langle { {\partial^2 { V} }
                        \over
               {\partial {\phi'}^1_1 \partial {\phi'}^3_3}
             }
     \rangle
\nonumber \\ 
\left( M^2_\eta \right)_{33} &=& 
-  { {\alpha_1 m_\pi^2 m_{\pi'}^2} \over {2 
A_1} }
+ 2
     \langle { {\partial^2 { V} }
                        \over
               {(\partial {\phi'}^1_1)^2}
             }
     \rangle
\nonumber \\ 
\left( M^2_\eta \right)_{34} &=& 
\sqrt{2}
     \langle { {\partial^2 { V} }
                        \over
               {\partial {\phi'}^1_1 \partial {\phi'}^3_3}
             }
     \rangle
\nonumber \\ 
\left( M^2_\eta \right)_{44} &=& 
     \langle { {\partial^2 { V} }
                        \over
               {\partial {\phi'}^3_3 \partial 
{\phi'}^3_3}
             }
     \rangle
\label{big4by4}
\end{eqnarray}

The four new quantities are $c_3$, discussed earlier, and the
``bare" primed squared masses:
\begin{equation}
 \langle { {\partial^2 { V} }
                        \over
               {(\partial {\phi'}^1_1)^2}
             }
     \rangle,           \hskip 1cm                   
 \langle { {\partial^2 { V} }
                        \over
               {\partial {\phi'}^1_1 \partial {\phi'}^3_3}
             }
     \rangle, \hskip 1cm
                   \langle { {\partial^2 { V} }
                        \over
               {\partial {\phi'}^3_3 \partial
{\phi'}^3_3}
             }
     \rangle.               
\nonumber \\
\end{equation}
These four quantities may be found by inputing the masses of four
isosinglet pseudoscalars. The net result is that all four systems
discussed will be completely described if all the experimental masses
and the decay constants, $F_\pi,F_K$
are specified together with an assumed value for $x_\pi$.

\section{Comparison with experiment and discussion}

    In the preceding section we gave the tree level 
formulas resulting
from the $M-M'$ model with any SU(3)$_L \times$ SU(3)$_R 
\times$
U(1)$_A$ invariant terms together with
a single ``instanton" type term which mocks up the 
U(1)$_A$
anomaly and the simplest structure which mocks up the
 quark mass terms. Isotopic spin invariance was also assumed.
Information is provided for only the pseudoscalar nonets
and the strange scalar particles. Information about
the scalar isotriplets can be obtained by including isospin
violation effects while information about the scalar
isosinglets requires either assuming some specific form
for the invariant interaction terms or computing other
 physical quantities. These will be discussed elsewhere.
   Now we will input the experimental masses to try to
learn what the model has to say about the quark structure
of the various mesons being described. In particular we are
 interested in the mixing angles like $\theta_\pi$,
governing admixtures of $q{\bar q}$ and $qq{\bar q}{\bar q}$
in the physical states and the four quark ``condensate"
strengths, $\beta_a$ which are associated with this mixing
 in the present model.                       

    The well known lowest pseudoscalar nonet masses and decay
constants will be taken, for definiteness (considering the ambiguity
as to which member of a non trivial isospin multiplet to choose),
 to be:
\begin{eqnarray}
m_\pi = 0.137 \, {\rm GeV} &,& \, m_K = 0.496 \, {\rm 
GeV},  \nonumber \\
m_\eta = 0.548 \,  {\rm GeV}&,& \, m_{\eta^\prime} = 
0.958 \, {\rm GeV},
\nonumber \\
F_\pi  =  0.131\,  {\rm GeV}&,&   \ F_K  =  0.160 \, {\rm 
GeV} .
\label{inputs}
\end{eqnarray}                                                                 

    Next, let us consider what are the suitable experimental
inputs for the masses of the excited mesons, $\pi',K'.{\kappa}'$
and for the $\kappa$ meson itself. In the latest Review of particle
properties \cite{rpp} there are two dotted (i.e. considered 
established) candidates for excited pions below 2 GeV: the $\pi(1300)$
and the $\pi(1800)$. These particles could have four quark
components and/or radially excited two quark components.
 In fact, judging
from an investigation of excited baryons \cite{baryons}, it is likely
that both types are present. Clearly, however, for our present  
investigation it seems reasonable to 
assume that the four quark component is the dominant one
and to 
choose the lower mass object
as the more suitable one. Similarly there are two undotted (non 
established)
excited kaon candidates: the K(1460) and the K(1830). We will again
choose the lower value. As candidates for an excited strange scalar
there is a dotted $K_0^*(1430)$ and an undotted $K_0^*(1950)$
and we again choose the lower value. In the case of the low mass
strange scalar there is an undotted  $K_0^*(800)$ candidate,
which we will interpret, with the help of \cite{BFSS1}, to be closer
 to 900 MeV. We summarize these choices:
\begin{eqnarray}
m_{\pi'} = 1.30 \, {\rm GeV} &,& \, m_{K'} = \, 1.46 
\, {\rm 
GeV},  \nonumber \\
m_\kappa = 0.90 \, {\rm GeV}&,& \, m_{\kappa^\prime} = 
1.42 \, {\rm GeV}.
\label{excitedinputs}
\end{eqnarray}                                                       

    For the excited $\eta$ type pseudoscalar particles the Review
of particle properties lists, below 2 GeV, the possible masses
(all in GeV):  
\begin{equation}
1.294,\hskip1cm 1.410, \hskip1cm 1.476, \hskip1cm 1.760.
\label{excitedetas}
\end{equation} 
The first three of these are dotted but the fourth is undotted.
Here it seems more difficult to a priori choose which are
most relevant 
so we shall study all possible pairings in a systematic way.

    First let us discuss the $\pi-\pi'$, $K-K'$ and $\kappa-
\kappa'$ systems. After using the inputs of Eq. (\ref{inputs}),
all features of these systems in our model will,
as already discussed,  be determined  
 by specifying $x_\pi$. Table \ref{systpara}
shows the predicted physical parameters for three values of
$x_\pi$. For orientation we note that in the chiral model with
a single field, M  one has
\begin{eqnarray}
   \alpha_1 \rightarrow F_{\pi}/2=0.0655 \,  {\rm GeV} 
&,& \alpha_3 \rightarrow
F_K-\alpha_1=0.0945 \, {\rm GeV},
\nonumber \\
     A_1 \rightarrow \frac{\alpha_1}{2}m_\pi^2=6.15 
\times 10^{-4} \,
{\rm GeV} ^3&,&
A_3 \rightarrow \frac{F_K}{2}m_K^2=0.01866 \, {\rm 
GeV}^3,
\nonumber \\
\beta_1 \rightarrow 0 &,& \beta_2 \rightarrow 0.
\label{oneMlimit}
\end{eqnarray}
The single M model corresponds to the choice $x_\pi=m_\pi^2$.
Increasing $x_\pi$ has the effect of increasing the admixture of
the ``four quark" field component in the physical pion.
The ``quark mass ratio", $A_3/A_1$ = 30.3
in the single M model is not very different
from the value of 31.2 obtained using the values in the $x_\pi=
0.019 \, {\rm Gev}^2$ column. The ${\bar q}q$ meson 
condensates
$\alpha_1$ and $\alpha_3$ are also very similar. Of course
the ``four quark" meson condensates $\beta_1$
amd $\beta_2$ are zero without $M'$. Despite the similarities,
the $6.4^o$ mixing angle already corresponds to about an 11
percent ``four quark" admixture in the physical pion 
wave function.
Considering that the accuracy of current algebra predictions
for low energy pion physics is roughly ten percent, it
seems that this choice of $x_\pi$ is the most plausible one. 
One sees from the second and third columns
that relatively small increases in $x_\pi$ lead to
 large increases in four quark admixture for the pion and
 the kaon. Interestingly, the behavior of the four quark 
admixture in the strange scalar meson $\kappa$ is quite different. 
When the pseudoscalars are closer to pure ``two quark" states
in the model the scalar has a large four quark admixture
($34.1^o$, with the choice
of $x_\pi$ in the first column). Thus the result is 
consistent with having a fairly large four quark component
 in the light scalars.

\begin{table}[htbp]
\begin{center}
\begin{tabular}{c||c|c|c}
\hline \hline 
& $x_\pi$=0.019\,  (GeV$^2$)&  
$x_\pi$=0.021 (GeV$^2$)& 
$x_\pi$=0.022 (GeV$^2$)
\\                                                 
\hline
\hline
$\theta_\pi$ (deg.) & $ - 6.4$
    & $- 19.1$
    & $- 22.7$
\\
\hline
$\theta_k$ (deg.)&
$- 11.2$    &
$- 22.9$    &
$- 26.2$
\\
\hline
$\theta_\kappa$ (deg.) &
$34.1$    &
$28.1$    &
$26.5$
\\
\hline
$A_1 ({\rm GeV}^3)$ &
$6.19 \times 10^{-4}$    &
$6.51 \times 10^{-4}$   &
$6.66 \times 10^{-4}$    
\\
\hline
$A_3 ({\rm GeV}^3)$ &
$1.94 \times 10^{-2}$    &
$2.07 \times 10^{-2}$    &
$2.12 \times 10^{-2}$    
\\
\hline
$\alpha_1$ (GeV) &
$6.51 \times 10^{-2}$    &
$6.20 \times 10^{-2}$    &
$6.06 \times 10^{-2}$
\\
\hline
$\alpha_3$ (GeV)&
$9.24 \times 10^{-2}$    &
$8.83 \times 10^{-2}$    &
$8.69 \times 10^{-2}$
\\
\hline
$\beta_1$ (GeV) &
$7.18 \times 10^{-3}$    &
$2.12 \times 10^{-2}$    &
$2.50 \times 10^{-2}$ 
\\
\hline
$\beta_3$  (GeV)&
$2.03 \times 10^{-2}$    &
$3.38 \times 10^{-2}$    &
$3.74 \times 10^{-2}$ 
\\
\hline
\hline
\end{tabular}
\end{center}
\caption[]{$\theta_\pi, \theta_K$ and $\theta_\kappa$ are 
respectively the ``four quark" admixtures in the $\pi, K$
and $\kappa$ states. $A_1, A_3$ represent the quark mass 
parameters while $\alpha_1, \alpha_3$ and $\beta_1, \beta_3$
represent respectively the two and four quark condensate
 strengths. These are plotted as functions of the assumed
``bare" pion squared mass, $x_\pi$.
}
\label{systpara}
\end{table}                                               

The analogs of the two quark condensates $\alpha_1=\alpha_2$ and
$\alpha_3$ are approximately equal, in agreement with the
usual assumption that
the vacuum is approximately SU(3) symmetric. The analogs of the
four quark condensates in this model are roughly an order of magnitude
smaller than the similarly normalized two quark condensates. They are
 furthermore seen to deviate appreciably from SU(3) symmetry. 
It should be noted, as discussed in ref, \cite{BFSS2} for example,
that the tensor indices for the primed mesons really correspond
to ``dual quark" or diquark indices in accordance with,
\begin{equation}
Q_a \sim \epsilon_{abc}{\bar q}^b{\bar q}^c.
\label{dualquark}
\end{equation} 
Thus in terms of the usual quarks,
\begin{equation}
\beta_1 \sim \langle{\bar d}d{\bar s}s\rangle,\hskip1cm 
\beta_2 \sim \langle{\bar u}u{\bar s}s\rangle,
\hskip1cm \beta_1 \sim \langle{\bar d}d{\bar u}u\rangle.
\label{interpretbetas}
\end{equation}
  
    Now consider the mixing of the four $\eta$ type 
fields in the model. The basis is given in Eq.(\ref{etabasis})
while the elements of the $4\times 4$ mass squared matrix
are given in Eq.(\ref{big4by4}).
The orthogonal transformation matrix, $K$ which relates
the mass eigenstate fields, $\Phi$ to the original ones is defined by 
\begin{equation}
\Phi_0=K \Phi.
\label{diagonalize}
\end{equation}
As discussed in the previous section, there are, after using 
the symmetry information, four new unknown parameters characterizing
the $\eta$ system. Thus taking the four mass eigenvalues
from experiment could in principle determine, together with results
from the $\pi-\pi'$, $K=K'$ and $\kappa-\kappa'$ systems,
everything about the $\eta$ system for a given value of $x_\pi$.
However there is no guarantee that there will be an exact solution
for all choices of experimental parameters. This is the case,
in fact, so we will search numerically for a 
choice of ``theoretical" masses which will best fit the
experimental inputs. The criterion for goodness of fit
 will be taken to be the smallness of the quantity: 
\begin{equation}
\chi \equiv \sum_{i} |m_i^{\rm exp.} - m_i^{\rm theo.}|\, / \,m_i^{\rm
exp.}.
\label{chi}
\end{equation}
As shown in Eq. (\ref{excitedetas}), there are three established
candidates and one
not yet established candidate below 2 GeV for the two excited
$\eta$ states. This yields six possible scenarios for choosing
them. The quantity $\chi$ for each choice is shown in
 Table \ref{etascenarios}  for three values of the parameter 
$x_\pi$. It may be observed that the fits typically get worse
with increasing $x_\pi$, so it is reasonable to consider
the choice 0.019 GeV${}^2$ for this quantity as we did
previously. The smallest values of $\chi$ are found for 
scenarios 5 and 6. However these both involve the $\eta(1760)$
state which is the one not yet established. The smallest
 value of $\chi$ using only established states
is scenario 2. This case corresponds to an exact fit
with eta type masses in GeV (experimental values in
parentheses for comparison):
\begin{eqnarray} 
&&      0.533 (0.548), \hskip2cm  0.963 (0.958),
\nonumber \\
&&      1.327 (1.294), \hskip2cm   1.716 (1.476).
\label{exactetas}
\end{eqnarray}

\begin{table}[htbp] 
\begin{center} 
\begin{tabular}{c||c|c|c} 
\hline \hline
Scenario & 
$x_\pi$=0.019 (GeV$^2$)& 
$x_\pi$=0.021 (GeV$^2$)& 
$x_\pi$=0.022 (GeV$^2$)
\\ 
\hline 
\hline
1:${\{}\eta(1295), \eta(1405){\}}$ & 
$6.23 \times 10^{-2}$ & 
$3.99 \times 10^{-1}$ &
$5.08 \times 10^{-1}$ 
\\ \hline 
2:${\{}\eta(1295), \eta(1475){\}}$ & 
$2.85 \times 10^{-2}$ & 
$3.39 \times 10^{-1}$ & 
$4.44 \times 10^{-1}$ 
\\
\hline
3:${\{}\eta(1295), \eta(1760){\}}$ &	
$2.35 \times 10^{-2}$ & 
$1.37 \times 10^{-1}$ & 
$2.28 \times 10^{-1}$ 
\\
\hline
4:${\{}\eta(1405),  \eta(1475){\}}$ &
$8.28 \times 10^{-2}$ & 
$3.63 \times 10^{-1}$ & 
$4.49 \times 10^{-1}$  
\\
\hline
5:${\{}\eta(1405), \eta(1760){\}}$ &	
$1.50 \times 10^{-2}$  &
$1.62 \times 10^{-1}$  &
$2.38 \times 10^{-1}$  
\\
\hline
6:${\{}\eta(1475), \eta(1760){\}}$ &	
$2.84 \times 10^{-2}$  &
$1.78 \times 10^{-1}$  &
$2.68 \times 10^{-1}$ 
\\
\hline
\hline
\end{tabular}
\end{center}
\caption[]{A goodness of fit quantity, 
$\chi \equiv \sum_{i} |m_i^{\rm exp.} - m_i^{\rm theo.}|\, / \,m_i^{\rm 
exp.}$, where the $m_i$ are the four mass eigenvalues of the $\eta$ type
fields, is given for 6 possible scenarios and for three values of
$x_\pi$. Each scenario corresponds to a choice of $\eta$ type fields
including the $\eta(548)$ and the $\eta(958)$ as well as the two
listed in the left hand column.}
\label{etascenarios}
\end{table}                                               

The detailed content of all the $\eta$ mass eigenstates
can be read off from the matrix $K^{-1}$. For scenario 2
we have,
\begin{equation}
K^{-1}=\left[ \begin{array}{c c c c}
               -0.570  & 0.750 & -0.023 & 0.333
\nonumber \\
               -0.329  & -0.573 & 0.142 & 0.737
\nonumber \\ 
		0.704 & 0.267 & -0.309 & 0.581
\nonumber \\ 
		0.267 & 0.192 & 0.940 & 0.088
                \end{array} \right] .
\label{Kinverse}
\end{equation}                                                 
Thus, in the present model there is an 89 percent
probability ($(K^{-1})_{11}^2 +(K^{-1})_{12}^2$) that
the $\eta(548)$ is a quark-antiquark state and
an eleven percent probability that it is a four quark 
state.
As expected, the $\eta(548)$ is most likely to be
in an ${\bar s}s$ state. In the case of the $\eta(958)$,
there is a 44 percent probability for it to be in a quark
antiquark state. There is a 54 percent probability for
it to be in the  four quark state ${\phi'}_3^3$. This situation has
some plausibility since in terms of ordinary quarks, the
latter state has the content ${\bar u}u{\bar d}d$ and
it should be most energetically favorable to bind a four
quark state made without strange quarks.

    The other scenarios which don't employ the
unconfirmed $\eta(1760)$ state (numbers
1 and 4) have contents very similar to the one in
Eq. (\ref{Kinverse}). On the other hand the three
 scenarios employing the $\eta(1760)$ have a 
rather different content, which seems unusual:
scenarios 3, 5 and 6 make the $\eta(958)$ almost
completely ${\phi'}^3_3$. 

    In scenario 2, which seems the most reasonable choice,
we notice that the $\eta(1295)$ has a 43 percent 
probability
of being in a four quark state while the $\eta(1475)$ has
an 89 percent probability of being in a four quark state.
To sum up, the value $x_\pi$ =0.019 GeV$^2$ leads to
fairly small four quark content in the light pseudoscalars-
$\pi, K, \eta$ at the same time that the light scalar $\kappa$
has an appreciable four quark component. The ``excited"
$\eta$'s are predominantly four quark states. The $\eta(960)$
is mainly two quark in content but has a non trivial four quark
piece. 

     The results obtained here provide supporting evidence for the 
feature, illustrated in the first treatment of this
 model \cite{BFMNS01},
that the lightest scalars, unlike the lightest pseudoscalars,
have appreciable four quark components. That model neglected
quark masses and used the simplified choice of terms shown 
 in Eq. (\ref{mixingpot}).
The more recent treatment of ref. \cite{nr},
includes two additional invariant terms
beyond those in Eq. (\ref{mixingpot})
(although not all the renormalizable terms 
shown in Appendix A) as well as four
types of quark mass splitting terms.
 Our results for the present treatment, where quark 
masses are included and which holds for any
 possible SU(3)$_L \times$ 
SU(3)$_R \times$
U(1)$_A$ conserving terms, are also in  
qualitative agreement for the $\pi$-type, K-type, $\eta$-type
and $\kappa$-type states with that treatment. 
Roughly, this may be expected
since the present approach includes any choice of invariant terms.
However, we only used here the single quark mass splitting 
term of Eq. (\ref{vsb}). Thus the results seem
qualitatively robust with respect to the treatment of the
mass splittings.
 
    An interesting feature of our model is the presence of
``four quark" condensates as signaled by the non-zero
values of the $\beta_a$. To make a rough estimate of what this
corresponds to in quark language we proceed as follows.
In ref. \cite{MS} it was pointed out that the mass formulas
of the single M linear sigma model could be transformed to the 
``current algebra" ones \cite{ca} by the replacements:
\begin{equation}
A_a=m_a \Lambda^2, \hskip1cm \alpha_a=-\frac{\langle{\bar 
q}_a 
q_a\rangle}
{2\Lambda^2},
\label{2vev}
\end{equation}
where the $m_a$ are the (``current" type) quark masses and
$\Lambda$ is the QCD scale factor. Taking $A_1$ = $6.19 
\times
10^{-4}\, {\rm GeV}^2$ from the left column of Table 
\ref{systpara} and $m_1
\approx$ 5 MeV we get $\Lambda \approx 0.35$ GeV (and 
$\langle{\bar q}_a q_a\rangle
\approx -0.016 \, {\rm GeV}^3$). In the case of the four 
quark condensate,
as one sees from the discussion in the Introduction, there are several 
ways to couple the four quarks together to make scalars. We
are assuming that one such way has been selected. For that case,
it is reasonable to expect, on dimensional grounds, that 
\begin{equation}
|\langle{\bar d} d {\bar s}s\rangle| \sim 
\Lambda^5 
\beta_1 \approx 4 \times 10^{-5}\,
{\rm GeV}^6.
\label{4vev}
\end{equation}

    In comparing the scalar masses with experiment 
there are expected to be, as discussed in the first four sections
of ref. \cite{BFMNS01}, non-negligible corrections 
due to the use of unitary models for the pseudoscalar-
pseudoscalar scattering based on this
Lagrangian. We plan to report on this elsewhere. This should also enable
 us to study the isosinglet scalar masses. For both isosinglet scalars
and pseudoscalars, the inclusion of possible glueball states is another 
interesting topic we plan to pursue. The additional symmetry
breaking terms like those in Eqs. (\ref{altanom})
 and (\ref{sbmixer}) seem also
to be worth investigating.

\section*{Acknowledgments}
\vskip -.5cm
We are happy to thank A. Abdel-Rehim, D. Black, M. Harada,
S. Moussa, S. Nasri and F. Sannino for many helpful
related discussions.
The work of A.H.F. has been supported by the 2004 
Crouse Grant from the School of Arts and Sciences, SUNY 
Institute of Technolgy.
The work of R.J. and J.S. is supported in part by the U. S. DOE under
Contract no. DE-FG-02-85ER 40231.

\appendix
\section{Renormalizable model}
    The twenty one SU(3)$\times$ SU(3) invariant 
renormalizable
terms in $V_0$ of Eq. (\ref{mixingLsMLag}) which can be made out of $M$ 
and $M'$ are:    
\begin{eqnarray}
V_0 =&-&c_2 \, {\rm Tr} (MM^{\dagger}) +{\tilde c}_3 \, ({\rm det} M + 
{\rm {\rm h.c.}}) +
 c_4^a \, {\rm Tr} (MM^{\dagger}MM^{\dagger})
 + c_4^b \, \left( {\rm Tr} (MM^\dagger)\right)^2
\nonumber \\
 &+& d_2 \, 
{\rm Tr} (M^{\prime}M^{\prime\dagger})
{\rm Tr} (M M'^\dagger)
     + d_3 \, ( {\rm det} M' + {\rm {\rm h.c.}})
      + d_4^a \, {\rm Tr} (M' M'^ \dagger M' M'^ \dagger)
     + d_4^b \, \left( {\rm Tr} 
(M^{\prime}M^{{\prime}{\dagger}})\right)^2
\nonumber \\
     &+& e_2 \, ( {\rm Tr} (MM^{\prime\dagger})+ 
{\rm {\rm h.c.}})
\nonumber \\
     &+& e_3^a(\epsilon_{abc}\epsilon^{def}M^a_dM^b_eM'^c_f + 
{\rm h.c.}) +
  e_3^b(\epsilon_{abc}\epsilon^{def}M^a_dM'^b_eM'^c_f + 
{\rm h.c.}) 
\nonumber \\
     &+& e_4^a \, {\rm Tr} (MM^{\dagger}M^{\prime}M^{{\prime}{\dagger}})
      + e_4^b \, {\rm Tr} (MM^{{\prime}{\dagger}}M^{\prime}M^{\dagger})
\nonumber \\
     &+& e_4^c \, [{\rm 
Tr} (MM^{\prime\dagger}MM^{\prime\dagger})+ {\rm h.c.}]
     + e_4^d \, [{\rm Tr} (M M^\dagger M M'^\dagger) + 
{\rm h.c.}]
 + e_4^e \, [{\rm Tr} (M'M'^\dagger M'M^\dagger) + 
{\rm h.c.}]
\nonumber \\
     &+& e_4^f \, {\rm Tr} (M M^\dagger) {\rm Tr} (M'M'^\dagger)
      + e_4^g \, {\rm Tr} (M M'^\dagger) {\rm Tr} (M'M^\dagger)
      + e_4^h \, [( {\rm Tr} (M' M'^\dagger))^2 + 
{\rm h.c.}]
\nonumber \\ 
      &+& e_4^i \, [ {\rm Tr}(M M^\dagger){\rm Tr}(M 
M'^\dagger) + {\rm h.c.}]
      + e_4^j \, [ {\rm Tr} (M' M'^\dagger) {\rm Tr} (M'M^\dagger) + h. 
c.].        
\end{eqnarray}
\label{renormV0}
Notice that among these terms, those with the coefficients
$c_2, c_4^a, c_4^b, d_4^a, d_4^b, e_3^a, e_4^a, e_4^b, e_4^f, e_4^g$ and
$e_4^h$ are $U(1)_A$ invariant.            
    It also may be of some interest to write down the twenty one
renormalizable terms, linear in the matrix A, which transform like the 
QCD quark mass 
terms under SU(3)$\times$ SU(3).  
 Again, for this listing, the U(1)$_A$
transformation property of the mass terms in the fundamental QCD
Lagrangian is respected only for the terms with
the coefficients $k_1, k_3, k_4, k_9, k_{11}, k_{12}, k_{17}, k_{21}$.
                                                                        
\begin{eqnarray}
V_{SB} =&+&k_1[ {\rm Tr} (AM)+ {\rm h.c.}]+k_2[ {\rm Tr} 
(AM')+ {\rm h.c.}]
\nonumber \\
      &+&k_3[ {\rm Tr} (AMM^\dagger M) + {\rm h.c.}]+k_4[  
{\rm Tr} 
(AMM'^\dagger M')+ {\rm h.c.}]
\nonumber \\
      &+&k_5[ {\rm Tr} (AMM^\dagger M')+ 
{\rm h.c.}]+k_6[{\rm Tr} 
(AMM'^\dagger M)+ {\rm h.c.}] 
\nonumber \\ 
&+&k_7[{\rm Tr} (AM'M'^\dagger M') + {\rm h.c.}]+k_8[ 
{\rm Tr} (AM'M^\dagger 
M)+ 
{\rm h.c.}]
\nonumber \\
&+&k_9[ {\rm Tr} (AM'M'^\dagger M)+ {\rm 
h.c.}]+k_{10}[{\rm Tr} 
(AM'M^\dagger M')+ {\rm h.c.}]
\nonumber \\
   &+&k_{11}[{\rm Tr} (AM)+ {\rm h.c.}]{\rm 
Tr}(MM^\dagger)
\nonumber \\
   &+&k_{12}[{\rm Tr} (AM)+ {\rm h.c.}]{\rm 
Tr}(M'M'^\dagger)
\nonumber \\
  &+&k_{13}[{\rm Tr} (AM) {\rm Tr}(MM'^\dagger) + 
{\rm h.c.}]+k_{14}[{\rm Tr}
(AM)
{\rm Tr}(M'M^\dagger) + {\rm h.c.}]
\nonumber \\
   &+&k_{15}[{\rm Tr} (AM')+ {\rm h.c.}]{\rm 
Tr}(MM^\dagger)
\nonumber \\
   &+&k_{16}[{\rm Tr} (AM')+ {\rm h.c.}]{\rm 
Tr}(M'M'^\dagger)
\nonumber \\
   &+&k_{17}[{\rm Tr} (AM'){\rm Tr}(MM'^\dagger) + 
{\rm h.c.}]+k_{18}[{\rm Tr}
(AM'){\rm Tr}(M'M^\dagger) + {\rm h.c.}]
\nonumber \\
   &+&k_{19} A_a^b \epsilon_{bcd}\epsilon^{aef} M_e^c M_f^d +
{\rm h.c.}
\nonumber \\
   &+&k_{20} A_a^b \epsilon_{bcd} \epsilon^{aef} {M'}_e^c 
{M'}_f^d + {\rm h.c.}
\nonumber \\
   &+&k_{21}A_a^b \epsilon_{bcd} \epsilon^{aef}M_e^c 
{M'}_f^d + {\rm h.c.}           
\label{renormVsb}
\end{eqnarray}

\section{Some isospin relations}

We give examples of relations which follow from isotopic
spin invariance:
\begin{eqnarray}
\langle \frac{\partial^2 V}{\partial \phi_2^2 \partial \phi_2^2}
\rangle &=&
\langle \frac{\partial^2 V}{\partial \phi_1^1 \partial \phi_1^1}
\rangle,
\nonumber \\
\langle \frac{\partial^2 V}{\partial \phi_2^2 \partial \phi_3^3}
\rangle &=&
\langle \frac{\partial^2 V}{\partial \phi_1^1 \partial \phi_3^3}
\rangle,
\nonumber \\                                                        
\langle \frac{\partial^2 V}{\partial \phi_1^2 \partial \phi_2^1}
\rangle &=&
\langle \frac{\partial^2 V}{\partial \phi_1^1 \partial \phi_1^1}
\rangle -
\langle \frac{\partial^2 V}{\partial \phi_1^1 \partial \phi_2^2}
\rangle.
\label{ispinrelations}                                                                            
\end{eqnarray}
Similar relations hold when V is differentiated with respect
to two primed fields and with respect to one primed
 and one unprimed field.

\end{document}